\documentclass[aps,prl,twocolumn,superscriptaddress,floatfix,showpacs,citeautoscript,longbibliography]{revtex4-2}

\usepackage{amsmath}
\usepackage{amssymb}
\usepackage{amsfonts}
\usepackage{dsfont}
\usepackage{graphicx,graphics}
\usepackage{color}
\usepackage[autostyle]{csquotes}
\usepackage{braket}
\usepackage[dvipsnames]{xcolor}
\renewcommand{\arraystretch}{1.8}
\usepackage[sort&compress]{natbib}
\usepackage{subfig}
\usepackage{float}
\usepackage{ragged2e}
\usepackage[capitalise,nameinlink]{cleveref}
\crefname{section}{App.}{Apps.}
\newcommand{\comment}[1]{}

\begin{document}

\title{Optomechanical Cooling without Residual Heating}
\author{Surangana Sengupta}
\affiliation{Institute for Complex Quantum Systems and IQST, Ulm University, Albert-Einstein-Allee 11, D-89069 Ulm, Germany}
\author{Bj\"orn Kubala}
\affiliation{Institute for Complex Quantum Systems and IQST, Ulm University, Albert-Einstein-Allee 11, D-89069 Ulm, Germany}
\affiliation{German Aerospace Center (DLR), Institute of Quantum Technologies, Wilhelm-Runge Straße 10, 89081, Ulm Germany}
\author{Joachim Ankerhold}
\affiliation{Institute for Complex Quantum Systems and IQST, Ulm University, Albert-Einstein-Allee 11, D-89069 Ulm, Germany}
\author{Ciprian Padurariu}
\affiliation{Institute for Complex Quantum Systems and IQST, Ulm University, Albert-Einstein-Allee 11, D-89069 Ulm, Germany}
\email{surangana.sen-gupta@uni-ulm.de}

\date{\today}
\begin{abstract}
Resolved-sideband cooling is a standard technique in cavity optomechanics enabling quantum control of mechanical motion, but its performance is ultimately limited by quantum backaction heating. This fundamental effect imposes a limit on the minimum achievable mechanical phonon number, establishing a finite-temperature floor regardless of the applied cooling strength. We generalize the semi-classical model for optomechanical cooling to describe universal cavity Hamiltonians incorporating both passive and active nonlinearities. As a concrete demonstration, we analyze the simplest circuit optomechanical system that implements a nonlinear drive via a Josephson junction. Our analysis reveals that this active nonlinear drive can eliminate the residual heating backaction, thereby comparing favorably with alternative optomechanical cooling schemes based on passive nonlinearities \cite{MetelmanExp2023}. By successfully overcoming the finite-temperature floor that limits conventional schemes, our method paves the way for unprecedented quantum control over mechanical systems and establishes the experimental viability of zero-heating optomechanical cooling.

\end{abstract}

\maketitle
\textit{Introduction}---
Optomechanical cooling is an important technology for high-precision experiments, enabling applications from building new mass sensors \cite{MassSensingSqueezedlaser,PDjorweEPMassSensing2019,LiUltraHighMassSensing2022}, to probing fundamental quantum effects in massive mechanical objects \cite{Oriol_cooling_2011,Hartmann_2015,Hartmann_Cat_states,ChuQAcoustic2017,ChuCatstate2023}. The conventional cooling scheme combines a strongly-driven harmonic cavity mode and a mechanical element that modulates the cavity frequency \cite{Braginsky_book,Kippenbergbasics2007}. Photons at the drive frequency scatter inelastically, exchanging energy with the mechanical mode, which results in the anti-Stokes sideband in the emission spectrum and reduces the mechanical phonon occupation \cite{KippenbergStokesAntiStokes2005,Clerk_PRL99,Wilson-Rae, Marquardt2009-qx, Metcalfe2014-sv,Clerk_PRL99,Wilson-Rae}. The theoretical description is well understood and, at weak optomechanical coupling, a semi-classical approach accurately captures the physics by directly relating the optomechanical damping to the cavity's shot noise spectrum \cite{Clerk_PRL99, Clerk_quantum_noise}.
Despite its successes, the conventional method is fundamentally limited by the unwanted backaction of the cavity on the mechanical mode \cite{RevModPhys.86.1391}. This backaction manifests as a residual heating effect that sets a non-zero floor for the lowest phonon occupation achievable via optomechanical cooling. While this residual heating can be minimized by increasing the cavity drive detuning \cite{Clerk_PRL99}, maintaining the required cavity occupation at high detuning demands a correspondingly increased driving power \cite{TeufelGSCOOL2011,Chan2011}. In current experiments, this driving power is ultimately constrained either by fundamental experimental limitations or, more commonly, by the intrinsic anharmonic behavior of the cavity mode.

In this Letter, we re-examine the minimum phonon number floor by analyzing the effects of an active nonlinearity, a capability realized by extending the semi-classical theory \cite{Clerk_PRL99} to describe cavities under arbitrary Hamiltonian control. Utilizing this general framework, we theoretically demonstrate that driving the cavity with a specific tunable, \textit{active} nonlinearity can completely eliminate the unwanted backaction effects. This suppression is due to squeezing generated within the cavity, allowing us to identify optimal cooling regimes and derive the exact conditions for zero residual heating. Our nonlinear scheme establishes the feasibility of cooling to phonon occupation orders of magnitude below unity, and it complements and generalizes prior work using squeezed light injected into \cite{Suppression_stokes_vitali,SqueezeTeufel2017,SqueezeCool24} or generated within \cite{Huang2009,Gan2019,Lau2020,SqueezelightcoolVitali,Duffing_Cooling2020,SqueezeCool24} the cavity, and recent investigations into passive Kerr nonlinearities \cite{MetelmanExp2023,Laflamme2011,Nation2008,MetelmanTheory2025,Deeg2025}. 

\textit{Model}--- We consider a generic optomechanical system where the mechanical degree of freedom $y$ modulates the cavity's resonance frequency $\omega_c(y)$. The driven cavity dynamics is governed by an arbitrary time-dependent Hamiltonian $\hat{H}_\text{cav}(t)$. This formulation naturally incorporates both active (e.g., a nonlinear drive with frequency $\omega_d$) and passive nonlinearities (e.g., a Kerr term \cite{MetelmanExp2023}).
Applying the rotating wave approximation (RWA), as is usual \cite{RevModPhys.86.1391}, yields the time-independent cavity Hamiltonian $\hat{H}_\text{cav}^\text{RWA}$, 
so that the total optomechanical Hamiltonian is $\hat{H}_\text{RWA}=\hat{H}_\text{cav}^\text{RWA}+\hat{H}_m+\hat{H}_\text{int}$.
The mechanical mode with frequency $\omega_m$ is governed by $\hat{H}_m=\hbar \omega_m (\hat{b}^\dagger \hat{b}+\frac{1}{2})$ and the optomechanical interaction follows as $\hat{H}_\text{int}=\hbar G\hat{y}\hat{a}^\dagger \hat{a}$,
obtained by linearizing $\omega_c$ \cite{Kippenbergbasics2007,RevModPhys.86.1391} with respect to the displacement $\hat{y}=y_\text{zpf}\left(\hat{b}^\dagger+\hat{b}\right)$. Here, $G=\partial \omega_c/\partial y$ is the single-photon optomechanical coupling, and $\hat{a}^\dagger$ is the cavity photon creation operator.
Including dissipation via input-output theory \cite{QNoise}, we arrive at the quantum Langevin equations,
\begin{align} \label{Eqn_a_dot}
\frac{d}{dt}{\hat{a}}= {}& \frac{i}{\hbar}\left[\hat{H}^\text{RWA}_\text{cav},\hat{a}\right]-\left(iG\hat{y}+\frac{\gamma}{2}\right) \hat{a} +\sqrt{\gamma}\hat{a}_\text{in}  \\
\frac{d}{dt}{\hat{b}}= {}& \left(-i\omega_m-\frac{\gamma_m}{2}\right) \hat{b} -iGy_\text{zpf}\hat{a}^\dagger \hat{a} +\sqrt{\gamma_m}\hat{b}_\text{in} \label{Eqn_b_dot}
\end{align}
where $\gamma$ is the cavity decay rate and $\gamma_m$ is the damping of the mechanical mode. The operators $\hat{a}_\text{in}$ and $\hat{b}_\text{in}$ model thermal noise \cite{QNoise}.
Under typical cryogenic conditions ($k_B T\ll\hbar\omega_c$), the cavity mode's thermal occupation is negligible ($\bar{n}_{c}^\text{T}\simeq 0$), while the mechanical mode remains highly occupied ($\bar{n}_m^\text{T}\gtrsim 10^2..10^3$) \cite{TeufelGSCOOL2011,MetelmanExp2023}.

\textit{Optomechanical damping rate: Preliminaries}--- We analyze the system under standard conditions: weak optomechanical coupling $g_0$ ($g_0= Gy_\text{zpf} \ll \omega_m, \gamma$) and high quality factors ($\omega_c/\gamma,\omega_m/\gamma_m\gg 1$, with $\gamma\gg\gamma_m$). The central quantity governing mechanical cooling is the optomechanical damping rate, $\Gamma_{\text{opt}}$. To leading order in the coupling $g_0$, it is determined by the asymmetry of the photon number spectrum $S_{\text{nn}}(\omega)$ around the mechanical frequency $\omega_m$ \cite{Clerk_PRL99,Clerk_quantum_noise}:
\begin{align} \label{Eqn_gamma_opt}
\Gamma_\text{opt} = g_0^2 [ S_{\text{nn}}(\omega_m)-S_{\text{nn}}(-\omega_m)],
\end{align}
with $S_{\text{nn}}(\omega)= \int^\infty_{-\infty} dt\ e^{i\omega t} \tilde{S}_{nn}(t) $ obtained from the photon number autocorrelation function $\tilde{S}_\text{nn}(t)=\langle \hat{a}^\dagger(t)\hat{a}(t)\hat{a}^\dagger(0)\hat{a}(0)\rangle-\langle \hat{a}^\dagger (t) \hat{a} (t)\rangle^2$ in the uncoupled limit ($g_0\to 0$). Starting from \cref{Eqn_a_dot}, our first objective is to obtain analytical expressions for these key quantities, while keeping the cavity Hamiltonian arbitrary.

Since achieving large damping rates requires driving the cavity to a high average photon occupation, $n=\braket{\hat{a}^\dagger\hat{a}} \gg 1$,  we can naturally employ a semi-classical approximation, which enables us to find a simple expression for the photon number spectrum, as detailed in the rest of this section. We decompose the cavity operator as $\hat{a} = \alpha_0 + \delta\hat{a}$, where $\alpha_0=\braket{\hat{a}} = A_0 e^{-i\theta_0}$ is the classical fixed point amplitude and $\delta\hat{a}$ is the quantum fluctuations operator. 
The fixed point $\alpha_0$ is the steady-state solution to the classical equation of motion $\partial_{\alpha^*}{\cal H}=i\hbar\gamma\alpha/2$, where the classical Hamiltonian ${\cal H}(\alpha,\alpha^*)$ is obtained from the normal-ordered $\hat{H}_\text{cav}^\text{RWA}$ by replacing the operators $\hat{a}$ ($\hat{a}^\dagger$) with $\alpha$ ($\alpha^*$) \cite{ Andrew_PRL,NoiseSwitch} (Appendix A).

The key quantity $\tilde{S}_\text{nn}(t)$ can then be expressed in terms of four elementary correlation functions, i.e. 
\begin{align}\label{Eqn_Snn_Si}
&\tilde{S}_\text{nn}(t)=  n\left[S_1(t)+S_2(t)+S_3(t)+S_4(t)\right].\\
&S_1(t)=e^{2i\theta_0}\braket{\delta \hat{a}(t)\delta \hat{a}(0)};\quad S_2(t)= \braket{\delta \hat{a}^\dagger(t)\delta \hat{a}(0)};\notag \\
&S_4(t)=e^{-2i\theta_0}\braket{\delta \hat{a}^\dagger(t)\delta \hat{a}^\dagger(0)};\quad S_3(t)=\braket{\delta \hat{a}(t)\delta \hat{a}^\dagger(0)}. \notag
\end{align}
\begin{figure*}[t] 
\includegraphics[width=2\columnwidth] {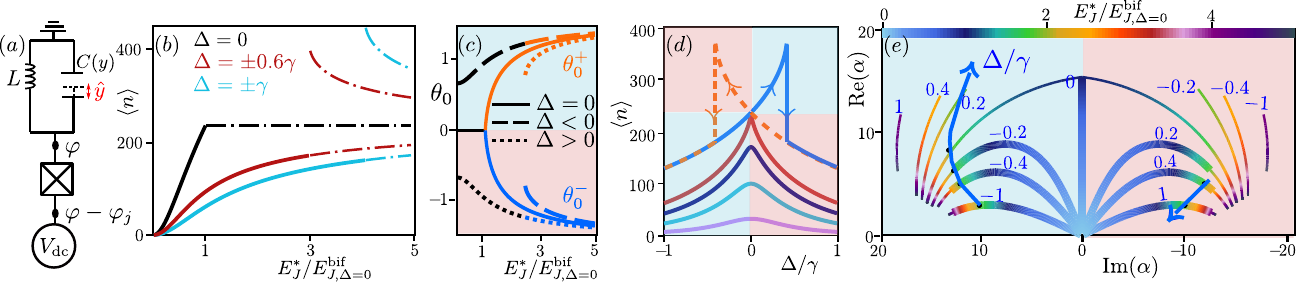}
\caption{ \justifying {\small (Color online.) 
Classical fixed points of the Josephson optomechanics circuit. 
(a) Sketch of the simplest circuit that implements a nonlinear drive. A dc-biased Josephson junction connects in series with a superconducting $LC$-circuit where the capacitance (or inductance) is modulated by a mechanical element. 
(b) Cavity photon number $n$ as function of driving strength $E_J^*$ for different detuning $\Delta$. The curves are symmetric for $\pm\Delta$. 
(c) Phase $\theta_0$ as function of driving strength $E_J^*$ for $\Delta=0,\pm0.4\gamma$. Monostable solutions (black) bifurcate into bistable solutions ($\theta_0^+$ orange and $\theta_0^-$ blue). In panels (c)-(e), blue and pink shading indicate the cooling and heating regimes. (d) Resonance curves $n(\Delta)$ for $E_j^*/\hbar\gamma=$ $100$ (purple), $200$ (light-blue), $300$ (navy), $404.40$ is $E_{J,\Delta=0}^\text{bif}/\hbar\gamma$ (dark-red), $750$ (blue/orange). (e) Classical fixed points in phase space, with $E_j^*$ indicated by the color scale. Thick lines represent monostable solutions; thin lines show bistable solutions. Curves correspond to different detunings $\Delta/\gamma$ (blue values). The blue arrow connects points (black) at $E_j^*/\hbar\gamma=750$, corresponding to the blue curve in (d). In all figures $\phi_0=0.06$.}}
\label{fig:Circuit_n_and_theta_Ej}
\end{figure*} 
These are derived straightforwardly from the semiclassical equation for fluctuations $\delta\hat{a}$, obtained by expanding Eq.~\eqref{Eqn_a_dot} to first order in $\delta\hat{a}/\alpha_0$, leading to a simple vector-matrix equation given in Appendix B. 
The correlation functions are found to depend on two universal parameters $\tilde{\Delta}$ and $r$, solely defined by local properties around the classical fixed points, with clear physical meaning: 
 the nonlinear effective detuning $\tilde{\Delta}=-(\partial_{\alpha\alpha^*}{\cal H})/\hbar$  and  the squeezing parameter $r=(-i/\hbar)e^{2i\theta_0}(\partial_{\alpha^*\alpha^*}{\cal H})$. Making use of the normal modes of fluctuations with characteristic eigenvalues $\lambda_{\pm}= -\frac{\gamma}{2} \pm \sqrt{|r|^2-\tilde{\Delta}^2}$, the solution for the correlators is readily determined, yielding the sought-after analytical results.
 
\textit{Optomechanical damping rate: Universal result}--- The photon number spectrum (\ref{Eqn_Snn_Si}) takes a remarkably compact form 
\begin{align}\label{Eqn_Snn_omega}
     \tilde{S}_\text{nn}(\omega)=\frac{n\gamma [(-\tilde{\Delta}+\omega+r_2)^2+(\gamma/2+r_1)^2]}{(\tilde{\Delta}^2-\omega^2+\gamma^2/4-|r|^2)^2+\gamma^2\omega^2}.
\end{align}
This quantity is the key ingredient that determines the optomechanical damping rate in Eq.~\eqref{Eqn_gamma_opt} as
\begin{align} \label{Eqn_gamma_opt_freq}
    \Gamma_\text{opt}(\omega)=\frac {4ng_0^2\gamma\omega (r_2-\tilde{\Delta})}{(\tilde{\Delta}^2-\omega^2+\gamma^2/4-|r|^2)^2+\gamma^2\omega^2},
\end{align}
and also determines the minimum phonon number (discussed below). We emphasize that these semiclassical expressions are applicable for arbitrary cavity Hamiltonians, and crucially, depend on the squeezing generated in the cavity by nonlinear effects. Specifically, $r_1$ and $r_2$ are the projections of the Hamiltonian squeezing parameter $r$ along the amplitude (radial in phase space) and phase (angular) directions, respectively, with $r=r_1+ir_2$. This universal relation between squeezing parameters, which are directly tunable in experiments, and the properties of optomechanical cooling, is the first central result of our work.

We illustrate the generality of \cref{Eqn_gamma_opt_freq} by reproducing previous results: for linear optomechanics, we recover the well-known expression \cite{Clerk_PRL99} by setting $\tilde{\Delta}=\Delta$ and $r=0$; for a linearly-driven Kerr cavity, we find $\tilde{\Delta}=\Delta+2{\cal K}n$ and $r=i{\cal K}n$ ($r_1=0$ and $r_2={\cal K}n$), which reproduces the results in Refs.~\cite{MetelmanExp2023} and \cite{MetelmanTheory2025}.
Having regained known results, we now go beyond and reveal the second central result of our work: squeezing generated by nonlinear cooling schemes can suppress unwanted residual heating, thereby removing the minimum phonon number floor. By eliminating this fundamental limitation of linear optomechanics (most detrimental for mechanical elements with low frequencies \cite{Clerk_PRL99,MetelmanTheory2025}), nonlinear schemes dramatically improve cooling for larger and slower mechanical elements.

\begin{figure}[t]
\includegraphics[width=1\columnwidth]{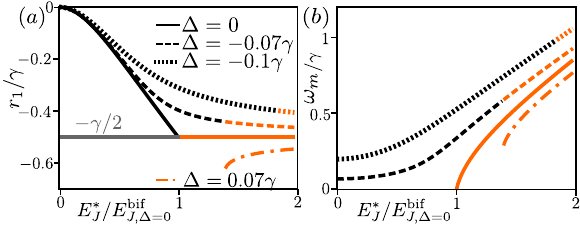}
 \caption{ \justifying{\small(Color online) Zero residual heating conditions for Josephson optomechanics. (a) Amplitude squeezing parameter $r_1$ versus normalized driving strength $E_J^*/E_{J,\Delta=0}^\text{bif}$ for specified detunings $\Delta$. Color indicates the monostable regime (black) and the bistable regime at $\theta_0^+$ (orange). (b) Mechanical frequency $\omega_m$ satisfying the minimal residual heating condition ($\omega_m=r_2-\tilde\Delta$), plotted against the driving strength, using the same parameters as (a).}}
 \label{fig:r1r2}
\end{figure}

\begin{figure}[t]
\includegraphics[width=1.02\columnwidth]{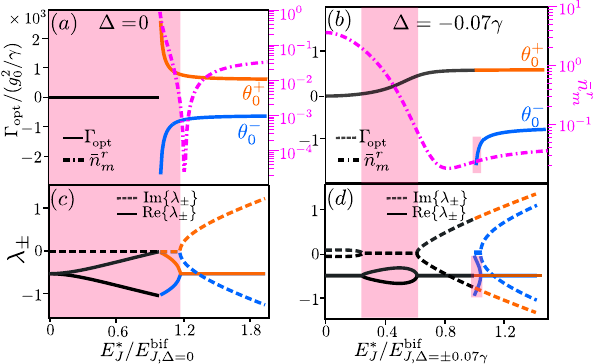}
 \caption{ \justifying{\small(Color online) 
 {Optomechanical damping and residual phonon number.} 
 (a,b) $\Gamma_\text{opt}$ and $\overline{n}_m^{\text{r}}$ at the optimal cooling frequency versus driving strength. Note the bifurcation of the monostable (black) solution into bistable solutions (orange and blue).
 (c,d) Complex eigenvalues $\lambda_\pm$ corresponding to the normal modes of fluctuations, shown for the same parameters as (a) and (b). The pink shading indicates regions where $\lambda_\pm$ are real, bounded by exceptional points. Detunings for (a,c) and (b,d) are indicated.} } 
 \label{fig:Panel_del0_del0.07}
\end{figure}

\textit{Minimum phonon number}--- In the cooling regime ($\Gamma_\text{opt}>0$), the optomechanical damping lowers the phonon occupation below its thermal value $\overline{n}_m^{\text{T}}$. The general expression for the minimum phonon number is given by \cite{Clerk_PRL99},
\begin{align} \label{Eqn_nm}
    \overline{n}_m = \frac{\Gamma_{\text{opt}}\overline{n}_m^{\text{r}}+\gamma_m \overline{n}_m^{\text{T}}}{\Gamma_{\text{opt}}+ \gamma_m},  
\end{align}
where $\overline{n}_m^{\text{r}}$ is the residual phonon occupation due to the unwanted backaction heating.  
It is quantified using detailed balance between the heating rate $g_0^2 S_\text{nn}(-\omega_m)$ and the cooling rate $g_0^2 S_\text{nn}(\omega_m)$, $\bar{n}_m^r=S_\text{nn}(-\omega_m)/\left(S_\text{nn}(\omega_m)-S_\text{nn}(-\omega_m)\right)$ which allows us to derive from \cref{Eqn_Snn_omega}
\begin{align} \label{Eqn_nmr_final_exp}
    \bar{n}_m^r{} =\frac{(\omega_m-[r_2-\tilde{\Delta}])^2+(\gamma/2+r_1)^2}{4[r_2-\tilde{\Delta}]\omega_m}.
\end{align} 
A remarkable consequence of the engineered nonlinearity is the possibility of vanishing residual heating ($\overline{n}_m^{\text{r}}=0$). This occurs precisely when the squeezing parameters satisfy $r_1=-\gamma/2$ and $\omega_m=r_2-\tilde{\Delta}$. The first condition requires fine-tuning $r_1$ to balance the damping and remove the detrimental heating noise [$S_\text{nn}(-\omega_m)=0$], while the second determines the frequency where this zero occurs. In the concrete example below [\cref{fig:Circuit_n_and_theta_Ej}(a)], $r_1=-\gamma/2$ is met at zero detuning ($\Delta=0$), or at finite detuning asymptotically at large driving amplitude (\cref{fig:r1r2}). The suppression mechanism is surprisingly effective: For approximate zeroes [$r_1\approx(-\gamma/2)$ and $\omega_m=r_2-\tilde{\Delta}$], the residual phonon number is drastically reduced compared to the established linear result \cite{Clerk_PRL99}, $\overline{n}_\text{m,lin}^{\text{r}}=(\gamma/4\omega_m)^2$, with the reduction given by $\overline{n}_m^{\text{r}}=\overline{n}_\text{m,lin}^{\text{r}}(1+2r_1/\gamma)^2$. Although our example system is not optimized for this effect, we find that a reduction of $\overline{n}_m^{\text{r}}$ by two orders of magnitude is feasible at finite detuning, achieving greater reductions at smaller detunings. The mechanical frequency that minimizes residual heating does not usually coincide with that for maximal $\Gamma_\text{opt}$. However, as shown in Figs.~\ref{fig:Panel_del0_del0.07}(a) and (b), a significant reduction of $\overline{n}_m^{\text{r}}$ below unity can still be realized at the frequency that maximizes $\Gamma_\text{opt}$, without fine tuning $r$. We further observe that connections reported for a Kerr cavity \cite{MetelmanTheory2025} between the suppression of $\overline{n}_m^{\text{r}}$, zeroes of $\Gamma_\text{opt}$ (defined by $r_2-\tilde{\Delta}=0$), and the exceptional points (EPs, defined by the $\lambda_\pm$ eigenvalue degeneracy $|r|^2-\tilde{\Delta}^2=0$), are found to be coincidental [see Figs.~\ref{fig:Panel_del0_del0.07}(c) and (d)]. Fundamentally, these three phenomena are triggered by \textit{independent} requirements.

The realization of zero residual heating by an engineered non-linearity is a central result of our work, that contrasts with linear and Kerr cavity optomechanics \cite{Clerk_PRL99,MetelmanTheory2025}. Its significance grows as the strength of cooling increases: in the ideal strong-cooling limit ($\Gamma_{\text{opt}} \gg \overline{n}_m^{\text{T}}\gamma_m$), the minimum phonon number approaches $\overline{n}_m \approx \overline{n}_m^{\text{r}} + \overline{n}_m^{\text{T}}(\gamma_m/\Gamma_{\text{opt}})$. Tuning $\overline{n}_m^{\text{r}}$ to zero removes the fundamental limitation on ground-state cooling, allowing $\overline{n}_m$ to reach levels orders of magnitude lower than previously possible, constrained only by the strength of the optomechanical damping. Since the squeezing parameters $r_1$ and $r_2$ can be fully controlled via an engineered non-linearity, the realization of zero residual heating could revolutionize optomechanical ground-state preparation.

We now illustrate our results with a concrete example based on a superconducting platform with active nonlinear driving using a Josephson junction (JJ), previously studied as a source of quantum states of light, including single-photons \cite{Rolland2019}, entangled photons \cite{Peugeot2021}, and photon multiplets \cite{Menard2022}. Note that bistabilities induced by nonlinearities of JJs are now routinely used for the read-out of superconducting quantum bits \cite{Siddiqi2004,Mallet2009}.

\textit{Josephson optomechanics}--- We consider the simplest circuit, schematized in Fig.~\ref{fig:Circuit_n_and_theta_Ej}(a). A dc-biased JJ efficiently drives the cavity via inelastic Cooper pair tunneling when the drive frequency $\omega_d=2eV_\text{dc}/\hbar$ (voltage bias $V_\text{dc}$) is near the cavity resonance $\omega_d\simeq\omega_\text{cav}$ \cite{Bright_Side,Gramich2013}.
Assuming a linear cavity mode with phase $\hat{\varphi}=\phi_0\left(\hat{a}^\dagger+\hat{a}\right)$, the total Hamiltonian is given by
\begin{align} \label{Eqn_Hamil_JP}
\hat{H}=\hbar\left(\omega_c + G \hat{y}\right) \hat{a}^\dagger \hat{a}+\hbar \omega_m \hat{b}^\dagger \hat{b}-E_J \cos (\hat{\varphi}+\omega_dt).
\end{align}
{\normalsize The non-linear Josephson driving (last term) contains the Josephson phase $\varphi_J=\varphi+\omega_\text{dc}t$, which follows from the circuit's Kirchhoff law [$V_\text{dc}+\hbar(\dot{\varphi}-\dot{\varphi}_J)=0$]. 
The RWA Hamiltonian (see Appendix C) is given by} 

\vspace*{-10pt}
\small
\begin{align} \label{Eqn_Hamil_RWA}
    \hat{H}_\text{cav}^\text{RWA}= {}& -\hbar  \Delta \hat{a}^\dagger \hat{a} + :\frac{ i E_J^*}{2}\left[\hat{a}^\dagger-\hat{a}\right]\frac{J_1(2\phi_0 \sqrt{\hat{a}^\dagger \hat{a}})}{\sqrt{\hat{a}^\dagger \hat{a}}}:
\end{align}
\normalsize
where $\Delta=\omega_{d}-\omega_c$ is the detuning and the renormalized Josephson energy $E_J^*=E_J e^{-\phi_0^2/2}$ acts as the effective driving strength. 
The colons denote normal ordering and $J_1$ the Bessel function.
 
The universal parameters are easily derived from the corresponding classical Hamiltonian $\mathcal{H}(\alpha, \alpha^*)$ as,
\begin{align}
{}& \tilde{\Delta} = \Delta-\frac{E_J^* \phi_0^2}{\hbar} J_1(2 \phi_0 A_0) \sin(\theta_0), \\
{}& r = -\frac{E_J^* \phi_0^2}{2\hbar} [J_{1}(2 \phi_0 A_0) e^{i\theta_0} + J_{3}(2 \phi_0 A_0) e^{-i\theta_0 }].
\end{align}
The governing equations for classical fixed points, $\alpha_0=A_0e^{-i\theta_0}$ are solved numerically in Figs.~\ref{fig:Circuit_n_and_theta_Ej}(b)-(e) (cf.~Appendix A). Two regimes have to be distinguished, one below and one above a bistability threshold $E_{J}^\text{bif}$.
At low driving $E_J^*$, the cavity occupation $n=A_0^2$ exhibits a standard Lorentzian response. However, above  $E_{J}^\text{bif}$ (which scales with detuning and $\phi_0^{-2}$ \cite{Andrew_PRL}), the solution bifurcates, leading to the hysteretic behavior seen in $n(\Delta)$ [Fig.~\ref{fig:Circuit_n_and_theta_Ej}(d)]. This hysteresis arises from the continuation of the monostable solution into the nearest bistable solution in phase space [Fig.~\ref{fig:Circuit_n_and_theta_Ej}(e)]. While phase symmetry holds at resonance, the symmetry is broken at finite detuning [Fig.~\ref{fig:Circuit_n_and_theta_Ej}(c)]. Crucially, strong optomechanical damping requires maximizing the magnitude of $n$. As shown in Fig.~\ref{fig:Circuit_n_and_theta_Ej}(b), $n$ increases with $E_J^*$ quadratically, then saturates at a value which scales as $\phi_0^{-2}$ \cite{Andrew_PRL}. 
Achieving the large cavity occupation necessary for efficient damping fundamentally requires designing a weak nonlinearity, achieved by engineering the cavity impedance $Z_\text{cav}$ such that $\phi_0=\sqrt{4\pi (Z_\text{cav}/R_K)} \ll 1$ ($R_K=h/e^2$ is the von Klitzing constant).

At resonance ($\Delta=0$), cooling ($\Gamma_\text{opt}>0$) is restricted to the bistable regime [\cref{fig:Panel_del0_del0.07}(a)], whereas finite negative detuning allows cooling both below and above bifurcation [\cref{fig:Panel_del0_del0.07}(b)]. The resulting minimum residual phonon number, $\overline{n}_m^r$, is greatly reduced, dropping well-below unity, with the strongest reduction occurring in the ideal ($\overline{n}_m^\text{r}=0$), resonant regime (where the pronounced dip in $\overline{n}_m^r$ corresponds to the coincidence of the optimal frequencies for $\Gamma_\text{opt}$ and $\overline{n}_m^\text{r}$). We now detail the distinct properties of cooling below and above bifurcation.

\begin{figure}[t]
\begin{center}
\includegraphics[width=1\columnwidth]{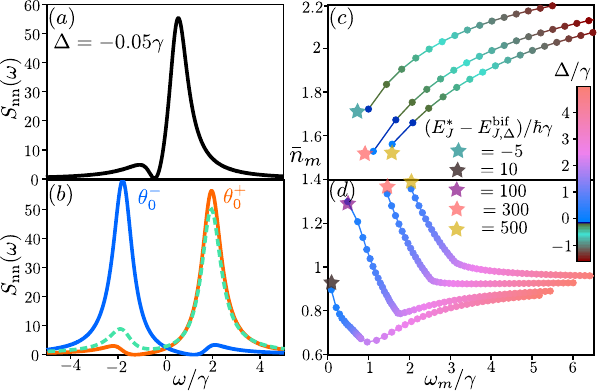}
\caption{ \justifying{\small (Color online.) 
{ Photon number spectrum and minimum phonon number.}  
Spectrum $S_\text{nn}$ at finite detuning, for the (a) monostable $E_J^*/E_{J,\Delta}^\text{bif}=0.92$, and (b) bistable $E_J^*/E_{J,\Delta}^\text{bif}=2.06$ cases, showing $S_\text{nn}(\theta_0^+)$ (orange) and $S_\text{nn}(\theta_0^-)$ (blue). The full quantum solution (green, dashed) fits to a weighted average $0.875 S_\text{nn}(\theta_0^+)+0.125 S_\text{nn}(\theta_0^-)$, biased towards $\theta_0^+$ for $\Delta<0$ [$\phi_0=0.2$ in (a) and (b)].
Minimum phonon number for various drive strengths and detunings with (c) negative detunings (cooling below and above bifurcation); (d) positive detunings (cooling only above bifurcation). The color scale indicates the detuning that maximizes $\Gamma_\text{opt}(\omega_m)$, while the distance to bifurcation, $(E_J^*-E_{J,\Delta}^\text{bif})/\hbar\gamma$, is kept fixed. Other parameters are given in \cref{tab:proposed_param}, Appendix D.}}
\label{fig:Snn_vs_omega}
\end{center}
\end{figure}

\textit{Cooling below bifurcation}--- In this regime, cooling ($\Gamma_\text{opt}>0$) occurs when the drive is red-detuned ($\Delta<0$), consistent with the slope of $n(\Delta)$ in Fig.~\ref{fig:Circuit_n_and_theta_Ej}(d) and similar to conventional optomechanics \cite{Clerk_PRL99}. While the nonlinearity causes $\Gamma_\text{opt}$ to undesirably saturate with increasing $E_J^*$, this is compensated by the concurrent suppression of the residual heating $\overline{n}_m^{\text{r}}$ [see the $S_\text{nn}$ dip in \cref{fig:Snn_vs_omega}(a)]. The resulting minimum phonon number $\overline{n}_m$ exhibits a more pronounced reduction at lower mechanical frequencies $\omega_m$ (Fig.~\ref{fig:Snn_vs_omega}(c), top curve). Specifically, utilizing parameters which align with the experiment of Ref.~\cite{MetelmanExp2023} (Table~\ref{tab:proposed_param}, Appendix D), the resulting value of $\overline{n}_m$ is an order of magnitude lower compared to that experiment. 

\textit{Ideal cooling above bifurcation}--- Above the bifurcation threshold, the system is bistable, possessing two stable solutions with positive $(A^+,\theta_0^+)$ and negative $(A^-,\theta_0^-)$ phases [see Figs.~\ref{fig:Circuit_n_and_theta_Ej}(c) and (e)]. Each solution has a distinct noise spectrum $S_\text{nn}$ [\cref{fig:Snn_vs_omega}(b)] and an associated damping rate $\Gamma_\text{opt}(\omega;\theta_0^\pm)$. Crucially, cooling always corresponds to the positive phase solution ($\theta_0^+$), regardless of the detuning sign, consistent with the slope of $n(\Delta)$ in the hysteresis region [\cref{fig:Circuit_n_and_theta_Ej}(d)]. The symmetry property $(A^+,\theta_0^+)|_\Delta=(A^-,\theta_0^-)|_{-\Delta}$ links the two solutions across detuning. To maximize cooling and achieve the lowest minimum phonon number $\overline{n}_m$ [Fig.~\ref{fig:Snn_vs_omega}(d)], it is advantageous to operate at the larger amplitude solution $A^+$, achieved when $\Delta>0$ [Fig.~\ref{fig:Circuit_n_and_theta_Ej}(e)].

The ideal cooling regime ($\overline{n}_m^\text{r}=0$) is found for cooling at resonance, in the $\theta_0^+$ state [Fig.~\ref{fig:Panel_del0_del0.07}(a)]. Here, cooling results from a spontaneous symmetry breaking in the damping mechanism. By symmetry at $\Delta=0$, the bistable damping rates are equal and opposite [$\Gamma_\text{opt}(\theta_0^+)=-\Gamma_\text{opt}(\theta_0^-)$]. Consequently, on long timescales that average over noise- and quantum-induced transitions, the net optomechanical damping vanishes ($\Gamma_\text{opt}^\text{avg}=0$). However, the symmetry is broken at intermediate times, shorter than the lifetime of bistable states. Within these times, we find ideal cooling conditions.

At finite detuning, the transitions between states are biased ($\Delta<0$ favors $\theta_0^+>0$), and the optomechanical damping rates become asymmetric [$\Gamma_\text{opt}(\theta_0^+) \neq -\Gamma_\text{opt}(\theta_0^-)$], as in Fig.~\ref{fig:Snn_vs_omega}(b). This results in a non-zero net optomechanical damping rate with a complex frequency dependence.

\textit{Conclusion} --- We have derived general expressions for optomechanical damping and residual heating that apply to arbitrary cavity Hamiltonians and experimental platforms, thereby pinpointing the central role of squeezing generated within the cavity. Our concrete example, optomechanics implemented in a simple Josephson circuit with strong nonlinear-driving and excellent tunability, demonstrates that control over a few universal parameters can both remove residual heating and maximize the net cooling rate. Our work establishes a compelling new direction for optimizing cavity nonlinearities to enhance optomechanical cooling.

\textit{Acknowledgements}---We gratefully acknowledge financial support from the IQST and the German Research Foundation (DFG) under AN336/13-1 and AN336/18-1.

\bibliography{journalabbreviations,references} 

\begin{onecolumngrid}
\section{End Matter}
\end{onecolumngrid}
\begin{twocolumngrid}
\textit{Appendix A: Classical cavity dynamics} --- 
\setcounter{equation}{0}
\renewcommand{\theequation}{A\arabic{equation}}
The classical equation of motion for the cavity dynamics in absence of coupling ($G\to 0$) can be elegantly derived from the input-output theory equation for $\hat{a}$,
 \begin{align} \label{Eqn_a_dot_Appendix_A}
     \dot{\hat{a}}={}&\frac{i}{\hbar}\left[\hat{H}^\text{RWA}_\text{cav},\hat{a}\right]-\frac{\gamma}{2}\hat{a}+\sqrt{\gamma}\hat{a}_\text{in}.
 \end{align}
We use the property of the commutator $\left[\hat{H}^\text{RWA}_\text{cav},\hat{a}\right]=[\hat{a}^\dagger,\hat{a}]\partial_{\hat{a}^\dagger}\hat{H}^\text{RWA}_\text{cav}$ and take the expectation value with respect to the cavity state, with the definition $\langle\hat{a}\rangle=\alpha$,
 \begin{align} 
     \dot{\alpha}={}&-\frac{i}{\hbar}\partial_{\alpha^*}{\cal{H}}-\frac{\gamma}{2}\alpha,
 \end{align}
 where $\cal{H}$ is obtained from the normal-ordered $\hat{H}^\text{RWA}_\text{cav}$ by replacing $\hat{a}(\hat{a}^\dagger)$ with $\alpha(\alpha^*)$.
 The classical fixed points $\alpha_0$ are steady-state solutions of $\partial_{\alpha^*}{\cal H}=i\hbar\gamma\alpha/2$.
 For the concrete example of Josephson optomechanics, we have
\begin{align}{\cal H}=-\hbar  \Delta |\alpha|^2 + \frac{ i E_J^*}{2}\left[\alpha^*-\alpha\right]\frac{J_1(2\phi_0 |\alpha|)}{|\alpha|}.
\end{align}
Setting $\alpha=Ae^{-i\theta}$, we obtain equations of motion for the average amplitude $A$ and phase $\theta$
 \begin{align} 
     \dot{A}={}&-\frac{\gamma}{2} A +\frac{E_J^* \phi_0}{2 \hbar} \cos (\theta) \left[J_0(2\phi_0A)+J_2(2\phi_0A)\right], \label{Eqn_A_dot}\\
    \dot{\theta}={}&-\Delta-\frac{E_J^* \phi_0}{2 A\hbar} \sin (\theta) \left[J_0(2\phi_0A)-J_2(2\phi_0A)\right].  \label{Eqn_theta_dot}  
 \end{align}
The steady state solutions $(A_0,\theta_0)$ are the classical fixed points presented in Fig.~\ref{fig:Circuit_n_and_theta_Ej}(b)-(e) of the main text. 

\textit{Appendix B: Photon number spectrum}--- 
\setcounter{equation}{0}
\renewcommand{\theequation}{B\arabic{equation}}
The dynamics of quantum fluctuations ($\delta \hat{a}$) around the classical fixed point $\alpha$ are derived by substituting $\hat{a}=\alpha+\delta\hat{a}$ into the input-output equation, Eq.~\eqref{Eqn_a_dot_Appendix_A}, and linearizing in $\delta \hat{a}/\alpha$. Linearizing the commutator is equivalent to keeping only the Hamiltonian terms quadratic in the fluctuation operators, leading to the fluctuations Hamiltonian $\hat{H}_\text{f}$
\begin{align}\label{Eqn_Hamil_Hf}
\hat{H}_\text{f} = (\partial_{\alpha\alpha^*}{\cal H})\delta \hat{a}^\dagger\delta \hat{a}+\frac{1}{2}[(\partial_{\alpha\alpha} {\cal H})(\delta \hat{a})^2+\text{h.c.}].
\end{align}
This results in the dynamical equation
\begin{align*}
    \delta \dot{\hat{a}}=\frac{i}{\hbar}\left[\hat{H}_\text{f},\delta\hat{a}\right]-\frac{\gamma}{2}\delta\hat{a}+\sqrt{\gamma}\hat{a}_\text{in},
\end{align*}
which can be expressed as the matrix vector equation

\vspace*{-10pt}
\small
\begin{align} \label{Eqn_da_dot}
\begin{pmatrix}
\delta \dot{\hat{a}} \\ \delta \dot{\hat{a}}^\dagger 
\end{pmatrix}= \begin{pmatrix}
i\tilde{\Delta}-\frac{\gamma}{2} & re^{-2i\theta_0} \\
r^*e^{2i\theta_0} & -i\tilde{\Delta}-\frac{\gamma}{2}
\end{pmatrix}
\begin{pmatrix}
\delta \hat{a} \\ \delta \hat{a}^\dagger
\end{pmatrix}+ \sqrt{\gamma}\begin{pmatrix}
{a}_\text{in} \\ {a}_\text{in}^\dagger
\end{pmatrix},
\end{align}
\normalsize
where $\tilde\Delta$ and $r$ are the effective detuning and squeezing parameter defined in the main text. Differentiation of the elementary correlation functions $S_i(t)$ and using \cref{Eqn_da_dot}, directly yields their equations of motion

\vspace*{-10pt}
\small
\begin{align}\label{Eqn_Si}
\frac{d}{dt}\begin{pmatrix}
{S}_{1(3)} \\ {S}_{2(4)}
\end{pmatrix}= \begin{pmatrix}
i\tilde{\Delta}-\gamma/2 & r \\
r^* & -i\tilde{\Delta}-\gamma/2
\end{pmatrix}
\begin{pmatrix}
S_{1(3)} \\ S_{2(4)}
\end{pmatrix}.
\end{align}
\normalsize

The initial conditions $S_i(0)$ correspond to the steady-state limit of correlators of equal-time fluctuations (e.g. $\braket{\delta a(t) \delta a(t)}$). Their equations are derived using \cref{Eqn_da_dot},

\vspace*{-10pt}
\small
\begin{align}
\frac{d}{dt}
\begin{pmatrix}
   e^{2i\theta_0} \braket{\delta a \delta a} \\
   \braket{\delta a^\dagger \delta a} \\
  e^{-2i\theta_0} \braket{\delta a^\dagger \delta a^\dagger}
\end{pmatrix} = \tilde{M} \begin{pmatrix}
 e^{2i\theta_0}   \braket{\delta a \delta a} \\
   \braket{\delta a^\dagger \delta a} \\
 e^{-2i\theta_0}  \braket{\delta a^\dagger \delta a^\dagger}
\end{pmatrix}+ \begin{pmatrix}
        r \\ 0 \\ r^*
    \end{pmatrix},
\end{align}
\normalsize
where \small 
$\tilde{M}=\begin{pmatrix}
    2i\tilde{\Delta}-\gamma &&  2r&& 0 \\
    r^*  && -\gamma && r \\
    0  &&2r^*&& -2i\tilde{\Delta}-\gamma\end{pmatrix}$.
\normalsize

In the steady-state ($\frac{d}{dt}=0$), this yields the matrix equation for the initial conditions $S_i(0)$
\begin{align}
0 = \tilde{M} \begin{pmatrix}
 S_1(0) \\
   S_2(0) \\
 S_4(0)
\end{pmatrix}+  \begin{pmatrix}
        r \\ 0 \\ r^*
    \end{pmatrix}.
\end{align}
Combined with the commutation relation $S_3(0)-S_2(0)=1$, we find the initial conditions
\begin{align}
{}& S_1(0)=\frac{r}{\gamma-2i\tilde{\Delta}} \left(1+2S_2(0)\right); \quad S_4(0)=S_1^*(0); \notag \\
{}& S_2(0)=\frac{1}{2} \frac{|r|^2}{\tilde{\Delta}^2-|r|^2+\frac{\gamma^2}{4}};\quad S_3(0)= 1+S_2(0).\notag
\end{align}
These, together with \cref{Eqn_Si}, lead straightforwardly to the analytical expressions for $S_i(t)$. 
Then, using \cref{Eqn_Snn_Si}, we find the photon number spectrum $S_\text{nn}(\omega)$ [\cref{Eqn_Snn_omega}] by Fourier transform.

\textit{Appendix C: Josephson photonics} --- 
\setcounter{equation}{0}
\renewcommand{\theequation}{C\arabic{equation}}
We derive the RWA cavity Hamiltonian (Eq.~\eqref{Eqn_Hamil_RWA} in the main text) starting from the total Hamiltonian (Eq.~\eqref{Eqn_Hamil_JP}), absent coupling ($G\to 0$). We apply the unitary transformation $\hat{U}(t)=e^{-i\hat{a}^\dagger \hat{a} \omega_dt}$ to transition to the rotating frame,
\begin{align}
         \hat{H}^{\text{rot}}_{\text{cav}} = -\hbar \Delta \hat{a}^\dagger\hat{a}-\frac{E_J}{2}\left(e^{i\omega_{\text{dc}}t}e^{i\zeta(t)}+ \text{h.c.} \right)
\end{align}
where, $\zeta(t)=\phi_0 \left(\hat{a}^\dagger e^{i\omega_{\text{dc}}t}+\hat{a}e^{-i\omega_{\text{dc}}t}\right)$.

We expand the exponential terms using the Baker-Campbell-Hausdorff formula and then discard all rapidly oscillating terms. The resulting RWA Hamiltonian is given by the infinite series

\vspace*{-10pt}
\small
\begin{align}
    \hat{H}^{\text{RWA}}_\text{cav}= - \hbar \Delta \hat{a}^\dagger\hat{a}  -\frac{E_J^*}{2} \left(\sum_{k=0}^\infty C_k (\hat{a}^\dagger)^k (\hat{a})^{k+1}+ \text{h.c.} \right)
\end{align}
\normalsize
where $C_k=i\frac{(-1)^{k} (\phi_0)^{2k+1}}{k!\  (k+1)!}$. Using the definition of the Bessel function $ J_\alpha(x)=\sum_{\kappa=0}^\infty \frac{(-1)^\kappa}{ \kappa ! \ (\kappa+\alpha) !} \left(\frac{x}{2}\right)^{2\kappa+\alpha}$, we recover the compact form of Eq.~\eqref{Eqn_Hamil_RWA} in the main text, with the effective driving strength $E_J^*=E_J e^{-\phi_0^2/2}$.

\textit{Appendix D: Experimental realization}---
\setcounter{equation}{0}
\renewcommand{\theequation}{D\arabic{equation}}
To exemplify the power of the nonlinear cooling scheme, we adopt the mechanical and optomechanical coupling parameters from the excellent recent work Ref.~\cite{MetelmanExp2023} (see Table~\ref{tab:proposed_param}) and focus our analysis on the regime below the bistability threshold, avoiding issues of stability.
\renewcommand{\arraystretch}{0.8}
\begin{table}[h]
    \centering
     \begin{tabular}{ |ll| } 
\hline
Parameters & Values  \\
\hline
Mechanical frequency  & $\omega_m=2\pi \times 302$ kHz \\
Mechanical linewidth & $\gamma_m=2\pi \times 0.5$ Hz \\
$LC$-cavity frequency & $\omega_c=2\pi \times 8$ GHz \\
$LC$-cavity linewidth & $\gamma=2\pi\times 3$ MHz \\
Optomechanical coupling & $g_0=2 \pi \times 2.1$ kHz \\
Driving strength  &      $E_J^*=31.32\  \mu$eV \\
$LC$-cavity ZPF &         $\phi_0=0.06$ \\
Detuning of the drive &\ $-\Delta=2\pi \times 30$ kHz \\
Optomechanical damping rate & $\Gamma_\text{opt}= 2 \pi \times 1282.39$ Hz \\
Backaction heating & $\bar{n}_m^r=0.075$ \\
Minimum Phonon number & $\bar{n}_m \ (\bar{n}_m^\text{T}=2778)= 1.15$ \\
\hline
\end{tabular}
   \caption{\justifying{\small{Minimum phonon number using Josephson optomechanics (mechanical and coupling parameters from Ref.~\cite{MetelmanExp2023}).}}}
    \label{tab:proposed_param}
\end{table}

Our cooling design strategy is threefold: (1) We minimize the cavity nonlinearity by adopting for the cavity-phase zero point fluctuations (ZPF) the value $\phi_0=0.06$ (corresponding to a circuit impedance of $Z_\text{res}\simeq 7.4\ \Omega$). This ensures a relatively large cavity photon number $n\simeq 100$, see \cref{fig:Circuit_n_and_theta_Ej}(b). (2) We fix the driving strength $E_J^*$ just below the bifurcation threshold $E_J^\text{bif}$. (3) We tune the detuning $\Delta$ to maximize $\Gamma_{\text{opt}}(\omega_m)$ using \cref{Eqn_gamma_opt_freq}.

This optimized strategy, detailed in \cref{fig:Snn_vs_omega}(c) and Table~\ref{tab:proposed_param}, reduces the minimum phonon number by roughly an order of magnitude compared to the original experiment \cite{MetelmanExp2023}. The core advantage lies in the resulting low residual phonon number, which permits dramatic improvements with only marginal setup enhancements. For example, a minor increase in the mechanical quality factor (corresponding to $\gamma_m^\prime=2\pi \times 0.302 $ Hz at the same mechanical frequency $\omega_m$) is sufficient to achieve sub-unity cooling, reaching $\bar{n}^\prime_m=0.73$.

\end{twocolumngrid}

\end{document}